\documentclass[twocolumn,showpacs]{revtex4}
\usepackage{epstopdf}
\usepackage{mathbbol}              
\usepackage{graphics,graphicx,epsfig,ulem}

\begin{document}

\title{Giant electro-optic effect using polarizable dark states.}
\author{A. K. Mohapatra$^{1}$, M. G. Bason$^1$, B. Butscher$^2$, K. J. Weatherill$^1$, and C. S. Adams$^1$}
\affiliation{1. Department of Physics, Durham University,
Rochester Building, South Road, Durham DH1 3LE, England.\\
2. Physikalisches Institut, Universit\"at Stuttgart, Pfaffenwaldring 57, 70569 Stuttgart, Germany.}

\begin{abstract}
{The electro-optic effect, where the refractive index of a medium is modified by an electric field, is of central importance in non-linear optics, laser technology, quantum optics and optical communications. In general, electro-optic coefficients are very weak and a medium with a giant electro-optic coefficient would have profound implications for non-linear optics, especially at the single photon level, enabling single photon entanglement and switching. Here we propose and demonstrate a giant electro-optic effect based on polarizable dark states. We demonstrate phase modulation of the light field in the dark state medium and measure an electro-optic coefficient that is more than 12 orders of magnitude larger than in other gases. This enormous Kerr non-linearity also creates the potential for precision electrometry and photon entanglement.}
\end{abstract}


\maketitle

In 1875 Kerr showed that the refractive index ($n_r$) of a medium can be changed by applying an electric field~\cite{kerr} according to
$$\Delta n_r=\lambda_0B_0E_0^2~,$$
where $\lambda_0$ is the wavelength of the light field, $E_0$ is the applied electric field and $B_0$ is the electro--optic Kerr coefficient. Subsequently, the Kerr effect, or quadratic electro--optic effect, and the related linear electro--optic effect have become widely used in photonic devices such as electro-optic modulators (EOMs)~\cite{yariv,boyd}. The ac Kerr effect where the electric field is produced by another light beam is the basis of Kerr lens mode--locking~\cite{spen91}, and has led to the development of femto and attosecond pulses~\cite{brab00}. Outside these successes, the wider applicability of the Kerr effect is limited by the fact that, in general, the Kerr non--linearity is very small. A larger non--linearity occurs close to a resonance, but at the expense of higher absorption of the signal light. A way around this problem is to use electromagnetically induced transparency (EIT)~\cite{boll91,kasa95,eit_review} where an additional light field, the coupling beam, renders a medium transparent on resonance. Enhanced ac Kerr non--linearities were predicted~\cite{schm96}, and have been studied in experiments on Bose Einstein condensates~\cite{hau99} and cold atoms~\cite{kang03}. However, such an EIT medium produces insufficient non--linearity to implement single photon non-linear optics~\cite{eit_review}. In addition, the potential to implement all--optical quantum computation using the ac Kerr effect~\cite{chua95} is limited by pulse distortion effects~\cite{shap06}, so a new Kerr mechanism based on interactions~\cite{fried05} is desirable.

In this paper, we demonstrate a giant dc electro-optic effect in an EIT medium by coupling to a highly excited Rydberg state which has a large polarizability. This renders the transmission through the medium highly sensitive to electric fields produced either externally or internally due to interparticle interactions. The Rydberg states have a polarizability that scales as the principal quantum number, $n^7$, and the interactions between Rydberg atoms scale with an even higher power  ($n^{11}$ for van der Waals interactions)~\cite{gallagher}.
These strong interactions lead to strongly correlated quantum states~\cite{luki01} and could be used to entangle atoms~\cite{jaks00} or photons~\cite{fried05}.  Non-linear effects due to strong interactions have been observed in a number of experiments on cold Rydberg atoms~\cite{tong04,sing04,lieb05,vogt06,heid07}. In our experiment, we show that the special properties of Rydberg atoms lead to a giant electro-optic effect ($> 10^{-6}$ m/V$^2$), which is $12$ orders of magnitude larger than in other gases~\cite{inba00} and $6$ orders of magnitude larger than the Kerr cell based on Nitrobenzene~\cite{optics}. This giant Kerr non-linearity opens up the possibility to devise a new class of single particle detectors, precision measurements in electrometry, and single photon entanglement.

\begin{figure}[]
\begin{center}
\includegraphics[width=3.3in]{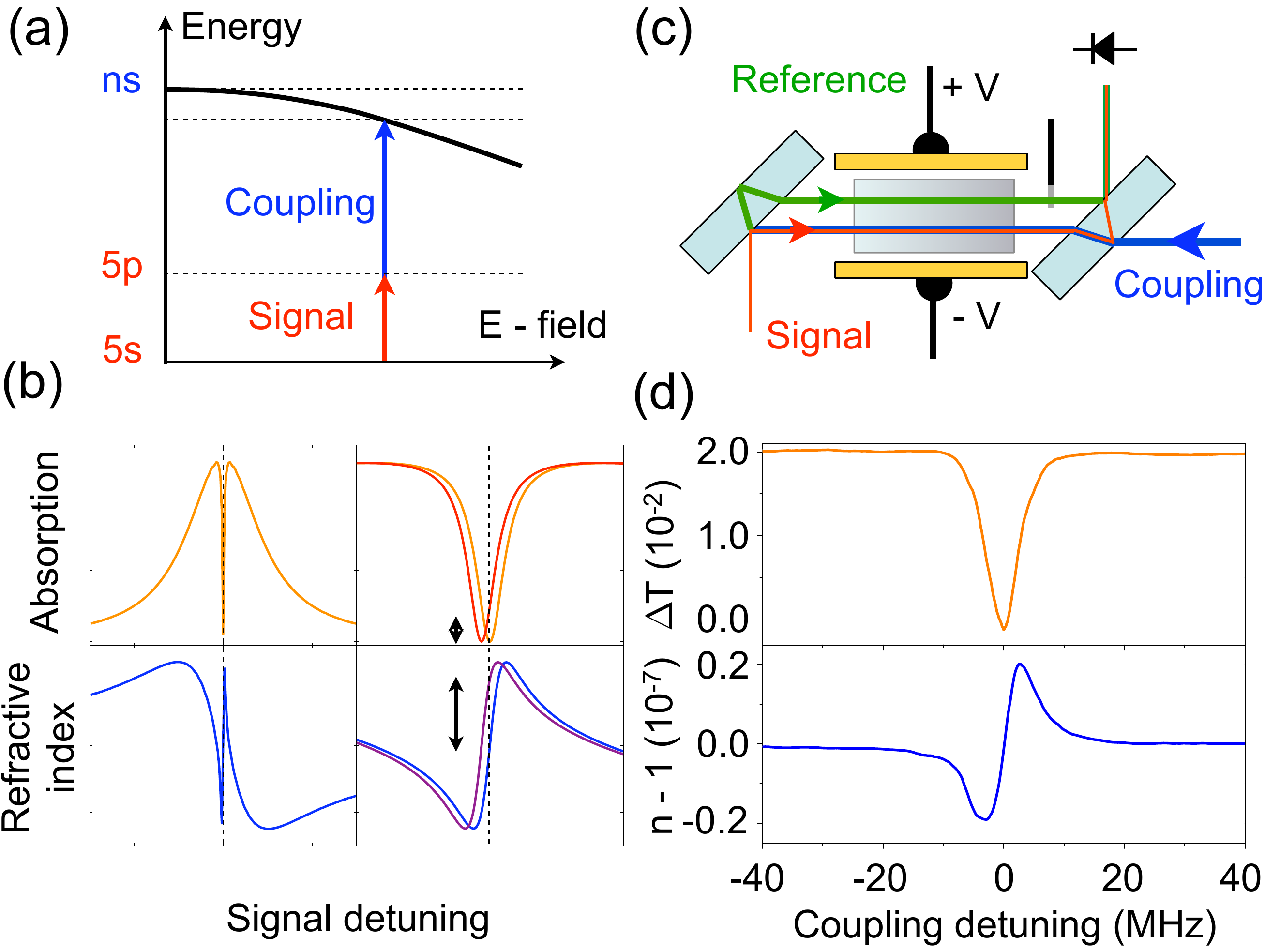}
\caption[]{(a) Schematic of the energy level diagram for Rydberg EIT. In contrast to the low lying states, the energy of the Rydberg state is extremely sensitive to electric fields. (b) The absorption and refractive index of the EIT medium as a function of the signal beam detuning. The images on the right show an expanded view of this transparency window. If the energy of the Rydberg state changes due to an applied electric field, the transparency window shifts leading to changes in the transmission and the refractive index. (c) The experimental set-up used to measure the electro-optic effect in a rubidium vapour cell. (d) The measured transmission and refractive index of the signal beam for room temperature atoms as the coupling beam is tuned through resonance with the Rydberg transition.}
\end{center}
\label{fig1}
\end{figure}

Recently, EIT involving a Rydberg level has been observed in thermal rubidium vapor~\cite{moha07} and a strontium atomic beam~\cite{maug07}. In this experiment, we employ Rydberg EIT to measure the electro-optic effect in rubidium vapor. A signal beam propagating through the rubidium vapor is  resonant with the 5s-5p transition, see Fig.~1(a). A counter-propagating coupling beam resonant with a Rydberg transition, 5p-ns (with $n=28-48$), drives the medium into a polarizable dark state~\cite{moha07}. For such a three level system, the absorption and the refractive index calculated for atoms at rest are shown in Fig. 1(b). When both the signal and coupling beams are exactly on resonance, the medium is prepared in a dark state corresponding to a superposition of the ground and Rydberg states~\cite{moha07}. In this dark state, the medium is transparent to the signal beam. If the energy of the Rydberg state is changed, this modifies the absorption and the refractive index (or dispersion) as indicated by the arrows in Fig. 1(b). The energy of the Rydberg state is highly sensitive to electric field due to its large polarizability as illustrated in Fig.~1(a). Thus, the refractive index of the dark state medium and hence, the phase of the signal beam, can be changed by applying an electric field. This induced phase modulation under the application of an ac electric field is measured using the set up shown in Fig~1(c). A weak laser beam (red) at $780.24$ nm, frequency stabilized~\cite{pear02} to the ${\rm s}~^2{\rm S}_{1/2}(F=3) \rightarrow 5{\rm p}~^2{\rm P}_{3/2}(F'=4)$ transition of $^{85}$Rb, is split into two paths (signal and reference) that propagate through a vapor cell. The signal and reference beams have the waist of 1 mm ($1/e^2$ radius) and power of $15~ \mu$W. The counter propagating coupling beam at $480$ nm (derived from a commercial frequency doubled diode laser system, Toptica TA-SHG) is overlapped with the signal beam. The 480~nm beam has a power of 140~mW with a waist of 0.8~mm ($1/{\rm e}^2$ radius) and is frequency stabilized to the $5{\rm p}~^2{\rm P}_{3/2}(F'=4) \rightarrow n{\rm s}~^2{\rm S}_{1/2}(F'')$ transition (the $n$s state hyperfine splitting is not resolved) using the Rydberg EIT resonance~\cite{moha07}. The signal and the reference beams are recombined using a second beam splitter, forming the Jamin interferometer that measures the relative phase. By blocking the reference beam, one can measure the transmission of the signal beam. The Rb vapor cell was placed inside a single layer magnetic shield and between two copper bar electrodes which were used to apply the electric field. Fig. 1(d) shows a typical measurement of the changes in the transmission and refractive index experienced by the signal beam as the coupling beam is tuned through the dark resonance for atoms at room temperature. As the signal and coupling beams are counter--propagating, there is a partial cancelation of the Doppler effect leading to a dark resonance line width of between 2 and 10 MHz depending on the laser power~\cite{moha07}.

Fig.~2 shows the response when the electric field is varied sinusoidally at a frequency close to 10~kHz and an amplitude of $\sim 2$~V/cm. With the reference beam blocked, we observe a modulation of the transmission with amplitude of about 1 $\%$, see Fig.~2(b). By unblocking the reference beam we observe the phase modulation of approximately $7$ mrad, see Fig.~2(c). The frequency spectrum of the amplitude and phase modulations are shown in Fig. 2(d) and (e). With the signal and coupling lasers locked just below the EIT resonance, the modulation in the transmission and the phase appear predominantly at the fourth and second harmonics of the modulation frequency, respectively. The Stark shift is proportional to $E^2$, so the refractive index (in the region of linear dispersion) is also proportional to $E^2$. This $E^2$ dependence is characteristic of the Kerr effect~\cite{yariv}. The transmission is modulated around the peak of the resonance and so is proportional to the energy shift squared and consequently to $E^4$, as observed. If the amplitude of the electric field is sufficiently large that the Stark shift is greater than the width of the dark resonance then the amplitude and phase modulation no longer have a simple power law dependence and higher harmonics are observed.

\begin{figure}[]
\begin{center}
\includegraphics[width=3.3in]{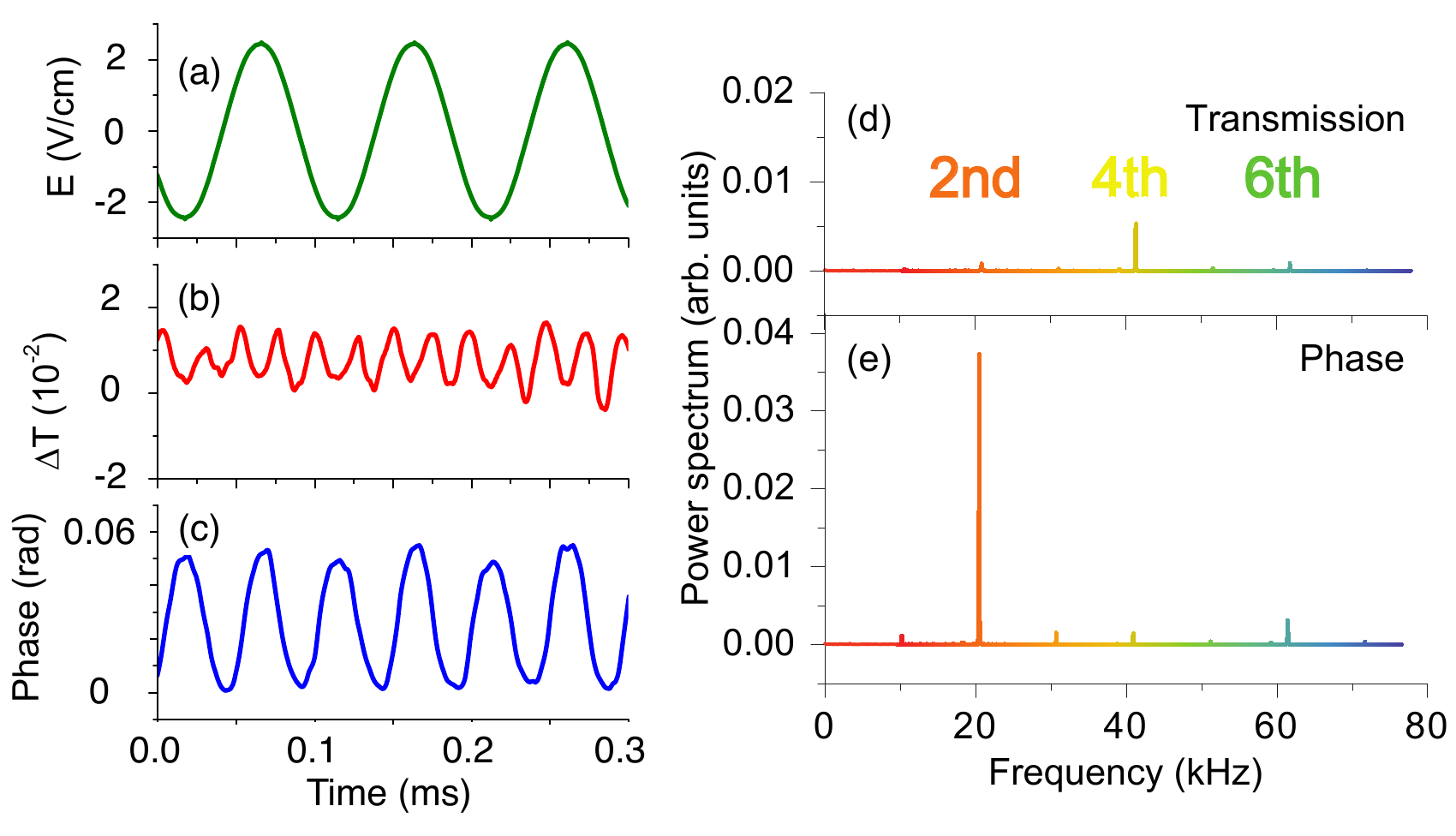}
\caption[]{The time dependence of (a) the applied electric field, $E$,  (b) the change in transmission through the medium, $\Delta T$ and (c) the phase of the signal beam for a room temperature Rb vapor cell. The Fourier transform of the (d) transmission and (e) phase data.}
\label{fig2}
\end{center}
\end{figure}

The magnitude of the phase shift in a room temperature cell as a function of the applied electric field ($E_0$) is shown in Fig. 3(a). The electric field dependence of the phase shift ($\Delta\phi$) follows the shape of the dispersion curve above resonance. By assuming the EIT resonance has a Lorentzian profile, the phase shift has the form, $\Delta \Phi=aE_0^2/(1+bE_{0}^4)$. The free parameters, $a$ and $b$ are determined from the fit and give the maximum phase shift. This allows one to calculate the wavelength independent Kerr coefficient, $B_0$ using $$\Delta\phi=2\pi B_0 \ell \vert E_{0m}\vert^2~,$$ where $\ell=0.075$~m is the length of the medium, and $E_{0m}$ is the electric field at which the maximum phase shift is observed. The measured Kerr coefficient for different principal quantum numbers, $n$ between $28$ and $48$ are presented in Fig 3. The experimental data show an $n^{*5.2}$ power law scaling, where $n^*=n-\delta$ and $\delta\approx 3.13$ is the quantum defect of the $s$ Rydberg series~\cite{li03}. The Kerr coefficient depends linearly on the change in refractive index of the medium. In the region of linear dispersion, this change is proportional to the product of the Stark shift and the slope of the dispersion feature. The Stark shift of the Rydberg state scales as $n^{*7}$. The slope of the dispersion signal is proportional to its amplitude and inversely proportional to its width. The amplitude has a quadratic dependence on the Rabi frequency of the $5$p-$n$s transition and scales as $n^{*-3}$. The width is power broadened in our experiment and reduces with higher $n^*$. This width scaling varies between $n^{*-1}$ and $n^{*-3/2}$ depending on the signal and coupling beam intensities. The observed power law dependence of the Kerr coefficient, $n^{*5.2}$, is consistent with these scalings.

The maximum measured Kerr coefficient at room temperature is of order $10^{-6}$~m/V$^2$ which is already much larger than in other media.
By heating the cell to $60^0$C, the Kerr coefficient increases by two orders of magnitude as shown in Fig. 3(b). Much larger Kerr coefficients are possible using a dense cloud of laser cooled atoms. For example, for a cold atom ensemble with diameter 10 microns at a density of $2\times 10^{12}$~cm$^{-3}$ and an $n=60$ dark state, one obtains a phase shift of $2\pi$ for a field of 3 V/m, corresponding to a Kerr coefficient of order $10^4$~m/V$^2$. In this high density regime it is necessary to avoid interaction effects \cite{tong04,sing04,lieb05,vogt06,heid07} by using a weak signal beam.

\begin{figure}[]
\begin{center}
\includegraphics[width=3.3in]{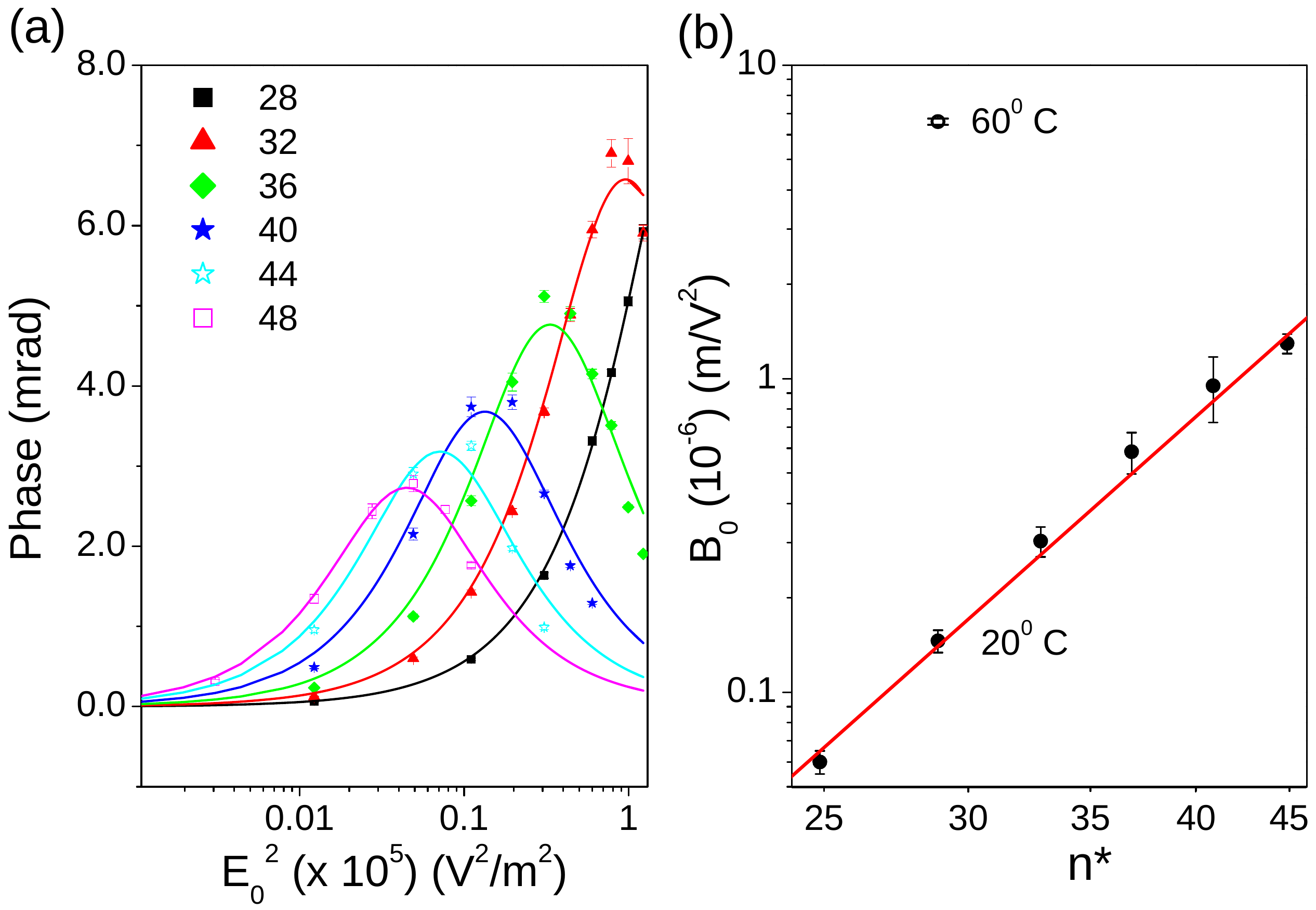}
\caption[]{(a) Measured phase shift as a function of the electric field amplitude with the coupling field tuned to an s Rydberg state with principal quantum number $n$ from 28 up to 48. (b) Measured Kerr constant in a vapor cell at $20^\circ$ C ($\bullet$) and $60^\circ$ C ($\circ$) as a function of the reduced principal quantum number, $n^*$. The solid line corresponds to a scaling of $n^{*\alpha}$, where $\alpha = 5.2\pm 0.2$.}
\end{center}
\label{fig3}
\end{figure}

A primary application of the electro-optic effect is the generation of light at a different frequency \cite{yariv,boyd}. For example, light propagating through a non-linear medium subject to an ac electric field acquires sidebands at harmonics of the modulation frequency. This sideband generation is easily observed using polarizable dark states. To measure the sidebands, we perform a heterodyne measurement on the signal beam. The reference beam of the Jamin interferometer shown in Fig 1(c) was blocked and the signal beam was interfered with another reference beam derived from the same laser but frequency shifted by $400$ MHz using an acousto-optic modulator. This 2$^{\rm nd}$ reference beam has a waist of 1~mm and power of 1~mW. The beat signal was detected using a fast photo-diode with a bandwidth of $1.2$ GHz. The power spectrum of the beat signal was recorded using a spectrum analyzer. Typical spectra are shown in Fig. 4. With the coupling laser turned off, we observe a single beat frequency. When the coupling laser is turned on, sidebands appear. The dominant sideband is the 2nd order as expected for the Kerr effect. The first harmonic sidebands can be observed by applying a DC field along with the modulating field. These data are recorded at a cell temperature of 60$^\circ$ C, where the atomic density and hence the phase modulation are significantly larger than at room temperature as shown in Fig. 3(b). With the signal and coupling lasers tuned to the 5s-5p-32s dark resonance, we observe sidebands with an intensity of $\sim1\%$ of the carrier for an applied voltage of $3$~V/cm. The sideband intensity is $60$ times larger than at room temperature, whereas the EIT peak height only increases by a factor of $2$. This shows that the observed sidebands are predominantly due to phase modulation.

The sideband amplitude as a function of modulation frequency, i.e., the modulation bandwidth (see inset of Fig. 4) is determined by the transient response of the EIT (dark state) resonance. If the coupling Rabi frequency is less than the decay rate of the intermediate state, $\gamma$, then the characteristic EIT response time is $\gamma$ \cite{li95,chen98}. Consequently for Rb atoms where the intermediate state lifetime is $\tau\sim 1/(2\pi\times6~{\rm MHz})$ one would expect a modulation bandwidth equal to the natural linewidth of $6$~MHz. The Lorentzian fit to the experimental data shown in the inset of the Fig. (4), has a full width half maximum of $10$~MHz. This value agrees with the resonance width obtained using Doppler free spectroscopy, and is slightly larger than the natural width due to pressure broadening from the background gas in the cell.

\begin{figure}[]
\begin{center}
\includegraphics[width=3.3in]{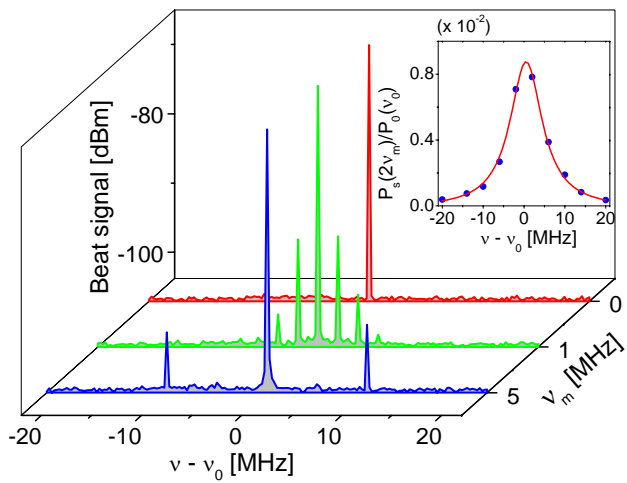}
\caption[]{The power spectrum of the light transmitted through the dark state ensemble with the coupling laser turned off is shown in red, where $\nu$ is the measurement frequency and $\nu_0$ is the carrier frequency. The sidebands generated with the coupling laser on and an electric field modulation with frequency $\nu_m$ equal to 1 and 5~MHz are shown in green and blue, respectively. The inset shows a Lorentzian fit to the power of the 2nd harmonic sidebands normalized to the power of the carrier. The full width half maximum of the fit is 10 MHz. The data were recorded in a Rb vapour cell at $60^0$ C.}
\end{center}
\label{fig4}
\end{figure}
\noindent

An interesting potential application of the giant electro-optic coefficient of polarizable dark states is precision electrometry. Such a device would provide the electric field analogue of precision atomic magnetometers \cite{shah07}. As measuring electric fields in free space is challenging, a dark state electrometer could find widespread applications in diverse areas such as geophysics and biophysics. For the predicted Kerr coefficient of a cold 10 micron ensemble, 10$^4$~m/V$^2$, a 1 mrad phase shift corresponds to a field of 400 $\mu$V/cm. In this case, a single charge at distance of 0.1~mm induces a phase shift of 100~mrad. Hence the electrometer could provide efficient detection of single charged particles. The advantage of using dark states to measure electric fields is the high bandwidth and ability to perform non-destructive measurements with a high repetition rate. Polarizable dark states can also be employed to detect single dipoles such as other Rydberg atoms which forms the basis of a proposed technique to entangle single photons in free space \cite{fried05}.

In summary, we have demonstrated a giant electro-optic effect in a gas based on polarizable dark states. The measured Kerr coefficient for thermal Rb vapour at room temperature, $B_0 \sim 10^{-6}$~m/V$^2$ is many orders of magnitude larger than in other electro--optic media. We have demonstrated phase modulation over a frequency range of 10 kHz to 10 MHz, and observed sidebands generation with an intensity up to 1$\%$ of the carrier. A significantly larger electro-optic coefficient could be obtained by increasing the optical depth for example by using laser cooled atoms. This giant electro-optic effect opens up the prospect of precision electrometry, single particle detection, and single photon entanglement.

We are grateful to E. Riis, T. Pfau, M. P. A. Jones, S. L. Cornish and I. G. Hughes for stimulating discussions, R. P. Abel for technical assistance and S. L. Cornish for loan of equipment. We thank the EPSRC for financial support.

\end{document}